# Optical Properties of Mono-Dispersed AlGaN Nanowires in the Single-Prong Growth Mechanism

*A. K. Sivadasan,*[1,*] *Avinash Patsha,*[1] *S. Polaki,*[1] *S. Amirthapandian,*[2] *Sandip Dhara,*[1,*]

*Anirban Bhattacharya,*[3] *B. K. Panigrahi,*[2] *and. A. K. Tyagi*[1]

[1]Surface and Nanoscience Division, Indira Gandhi Centre for Atomic Research, Kalpakkam-603 102, India

[2]Materials Physics Division, Indira Gandhi Centre for Atomic Research, Kalpakkam-603 102, India

[3] Institute of Radiophysics and Electronics, University of Calcutta, Kolkata-700009, India

**Corresponding Authors:** sivankondazhy@gmail.com, dhara@igcar.gov.in

*Abstract*

Growth of mono-dispersed AlGaN nanowires of ternary wurtzite phase is reported using chemical vapour deposition technique in the vapour-liquid-solid process. The role of distribution of Au catalyst nanoparticles on the size and the shape of AlGaN nanowires are discussed. These variations in the morphology of the nanowires are understood invoking Ostwald ripening of Au catalyst nanoparticles at high temperature followed by the effect of single and multi-prong growth mechanism. Energy-filtered transmission electron microscopy is used as an evidence for the presence of Al in the as-prepared samples. A significant blue shift of the band gap, in the absence of quantum confinement effect in the nanowires with diameter ~ 100 nm, is used as a supportive evidence for the AlGaN alloy formation. Polarized resonance Raman spectroscopy with strong electron-phonon coupling along with optical confinement due to the dielectric contrast of nanowire with respect to that of surrounding media are adopted to understand the crystalline orientation of a single nanowire in the sub-diffraction limit of ~100 nm using 325 nm wavelength, for the first time. The results are compared with the structural analysis using high resolution transmission microscopic study.

1. INTRODUCTION

The research towards wide–band gap group III–nitrides semiconducting material is very important because of its applications in blue to near UV light emitting devices. The main attractive material properties are enormous physical hardness, extremely large heterojunction offsets, high thermal conductivity, and elevated melting temperature.[1] Group III nitride materials are one of the most promising candidates for fabricating these short wavelength and high frequency optoelectronic devices.[2-4] InN, GaN, and AlN and its ternary (InGaN and AlGaN) and quaternary (InAlGaN) alloys are the well–known candidates for major optoelectronic applications.[3,5-7]

Scientific researchers all over the world already developed various 1D nanostructures such as nanotubes, nanowires (NWs), and nanocones for the binary AlN,[8-10] GaN,[11-14] and InN.[15-17] Because of the high aspect ratio, the NWs find enormous utility in various optoelectronics applications. The electromagnetic propagation will be very strong in NWs compared to other nanostructures and more than that, the high aspect ratio of NWs leads to the strain relaxation of the material without the formation of dislocation. The ternary alloy of $In_xGa_{1-x}N$ NWs were developed with a tunable band gap in the range of 1.1-3.4 eV, showing the emission from the near-IR to the near–UV region.[18] To attain the applications of complete electromagnetic spectral region, it is necessary to cover ranges again from blue to UV region. The prominent candidate for this wide bang gap material is ternary alloy of $Al_xGa_{1-x}N$ system with the tunable band gap at 3.4–6.2 eV.[19,20] $Al_xGa_{1-x}N$ is an ideal semiconducting candidate for the fabrication of such optoelectronic devices.[19-22] Now the research progressed up to the fabrication of single NW light emitting diode (LED) and photodetectors being optically coupled by waveguides.[21,23] The InGaN/AlGaN double–hetero structures are useful for making the blue

LEDs. It is possible to handle the entire electromagnetic spectrum ranging from UV to IR with the combined use of these two ternary alloy compositions.[1,3,5,24] However, for making such optoelectronic devices operating at short UV wavelengths, alloy compositions with greater Al content are required. The material properties of these materials lie on the border between conventional semiconductors and insulators, which add a degree of complexity to the development of efficient light–emitting devices.[25] In nanometer scale dimension, 1D AlGaN NWs used for demonstrating the quantum–wire–in–optical fiber (QWOF).[26] The quaternary alloy of InAlGaN materials are ideally suited for the realization of UV LEDs since they are applicable throughout the entire UV (200–400 nm) spectral range.[21,22]

Here, we report a simple technique of atmospheric pressure chemical vapor deposition (APCVD) for synthesizing AlGaN NWs in the catalyst assisted vapor-liquid-solid (VLS) process. In the present study, we mainly focus in the synthesis of AlGaN NWs of different size and shapes by regulating the distribution of Au catalyst nanoparticles (NPs) over the substrate. Role of Ostwald ripening is discussed in understanding the growth of AlGaN NWs via VLS mechanism. In order to confirm the presence of Al, we used the electron energy loss spectroscopy (EELS) and the energy–filtered transmission electron microscopic (EFTEM) studies. The AlGaN alloy confirmation is further confirmed by observing blue shift in the band gap in the absence of quantum confinement effect.

2. EXPERIMENTAL

AlGaN NWs were synthesized by APCVD technique in the VLS process. The Ga metal drop (99.999%, Alfa Aesar) was used as the Ga source. Ammonia (NH$_3$, 99.999%) with a flow rate 50 sccm was used as reactant gas. A thin film with an average Al metal thickness of 50 nm was used as the source for Al. The Au coated Si (100) substrates were used for the growth of AlGaN NWs in the VLS process. The Al thin film as well as Au islands were deposited on Si(100) substrates using a thermal evaporation method (12A4D, HINDHIVAC, India). In order to make various size of Au NPs, the Au islands were coated for 5 and 10 min with a constant evaporation rate (1Å /sec) and subsequent annealing at a temperature 900 $^{o}$C for 10 and 20 min in an inert (Ar) atmosphere, respectively. The Si substrate with Al thin film and the annealed Au NPs were placed separately in two high pure alumina boats (99.95%). The alumina boats along with the substrates were inserted into a quartz tube which was kept inside the two zone CVD furnace. The alumina boats were positioned in the tube, such a way that the Al thin film would be placed in the first and Au NPs will be in the second zone of the furnace shown later during discussion in the growth mechanism. Then furnace is programmed for temperatures of 1100 $^{o}$C and 900 $^{o}$C for Al thin film and Au NPs coated substrates, respectively. The temperature of the quartz tube was systematically increased to the growth temperature of 1100 $^{o}$C and 900 $^{o}$C for the respective zones with same ramp rate of 15 $^{o}$C min$^{-1}$. The NWs were synthesized for an optimized growth time of 120 min.

Morphological features of the as-prepared samples were analyzed using a field emission scanning electron microscope (FESEM, SUPRA 55 Zeiss). The structural and crystallographic nature of the AlGaN NWs were investigated with the help of a high resolution transmission electron microscopy (HRTEM, LIBRA 200FE Zeiss) by dispersing the NWs in isopropyl alcohol and transferred to Cu grids. EELS studies were carried out for identifying the presence of Ga, N, and Al in a single NW using in-column second order corrected omega energy filter–with an energy resolution of 0.7 eV. The distribution of the three elements Ga, N, and Al in NWs was studied by generating EFTEM corresponding to the core–loss energy of each element of the NWs. The vibrational properties of AlGaN NWs were studied using Raman spectroscopy (inVia, Renishaw,UK) with $Ar^+$ laser excitation of 514.5 nm and 1800 gr.mm$^{-1}$ grating used as a monochromatizer for the scattered waves. The thermoelectrically cooled CCD detector was used in the backscattering geometry. The spectra were collected using a 100X objective with numerical aperture value of 0.85. For understanding the band gap and optical quality of the material, the AlGaN NWs were excited with a UV laser of wavelength 325 nm and the photoluminescence (PL) spectra were recorded at room temperature using the Raman spectrometer with 2400 gr.mm$^{-1}$ grating and an adequate band pass filter.

## 3. RESULTS AND DISCUSSION

### 3.1 Morphological analysis

Morphologies of essentially two different types of NWs were investigated as shown in the figure 1. These differences in morphology were achieved by varying the number density distribution of Au NPs on the Si(100) substrate. Figure 1a shows the low magnification image of the NWs synthesized on Si(100) substrate with Au catalysts coated with an evaporation time

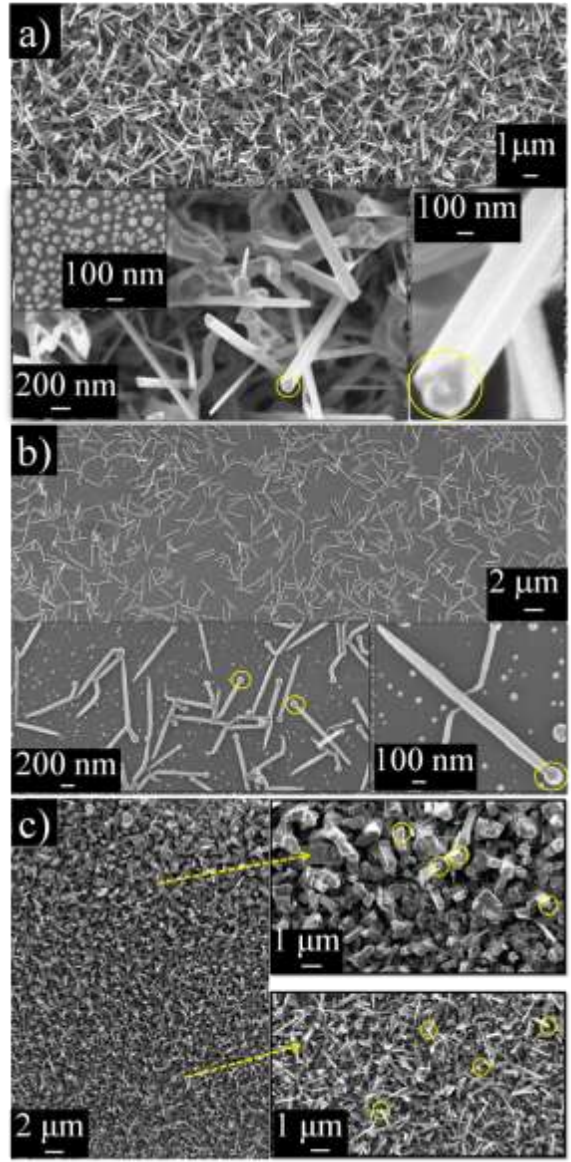

**Figure 1.** a) NWs synthesized on Si(100) substrate with Au catalysts coated with an evaporation time 10 min and subsequent annealing treatment. Outsets show high resolution images of the triangular shaped NW with Au NP at the tip of the NW as indicated by arrows. The size and shape of the Au NPs are also shown in the inset of one of the high resolution images. b) NWs synthesized with Au catalysts coated with an evaporation time 5 min and subsequent annealing treatment. Outsets show high resolution images of the cylindrically shaped NW with Au NP at the tip of the NW as indicated by circles. Along with these NWs, size of the Au NPs is visible in this high resolution image itself. c) Variation of the size distribution of NW in the same substrate from nearly cylindrical shape to triangular ones. Outsets show the high resolution images for the NWs from different regions of the substrate (as indicated by arrows) - triangular and cylindrically shaped NWs with Au NPs observed at the tip are indicated by circles.

10 min and subsequent annealing treatment as described in the experimental section. The dense and non–uniform size distributions of NWs are clearly observed all over the substrate. The high resolution image of the NWs with an average diameter of 50–150 nm is shown at the outsets of Figure 1a. A dense distribution of Au catalyst NPs with an average diameter 50–100 nm, which is used for the VLS growth of AlGaN NWs, is shown in the inset in one of the magnified FESEM images. It is obvious that almost all the Au catalyst NPs participated in the growth process leading to the dense distribution of NWs all over the Si(100) substrate. For the first sample, it is clearly visible that most of the NWs were grown with triangular shape in cross-section and also shown non–uniformity in the diameter of the NWs as it varies from 50–150 nm (outsets Fig. 1a). The high resolution FESEM images of the NWs show very smooth surface morphology with Au catalyst NP at the tip. The size of the triangular shaped NWs is higher than the size of the Au NPs observed at the tip. In the case of second sample (Fig. 1b), which was synthesized on Si(100) substrate with Au catalysts coated with an evaporation time 5 min, the distribution of NWs with almost circular cross section on the substrate less dense as compared to the first sample (Fig. 1a). Moreover a uniform size with an average diameter of 100($\pm$10) nm was also observed for the NWs. The mono-dispersed NWs with smooth morphology are spread all over the substrate and the size of the Au catalysts NPs are marginally higher than the diameter of the NWs (outsets of Fig. 1b). The Au NPs which took part in the VLS growth process of the NWs were having uniform size of ~120 nm and was found to be well separated. However, some of the Au NPs with lower diameters have not participated in the VLS growth process (outsets of Fig. 1b).

In order to understand the dependence of size and shape of NWs with respect to the arrangement of Au catalyst NPs over the substrate, we made a gradient of the distribution of Au NPs from low to dense on the same Si(100) substrate. This was achieved by the drop casting of uniform Au NPs (20 nm) dispersed in the water medium on the Si(100) and subsequent infra-red lamp heating to evaporate the water (not shown in figure). For achieving the gradient in the number density of Au NPs, the substrate is slightly tilted while drop casting. This Au NP drop casted Si(100) substrate was used for the VLS growth of NWs. The growth led to the formation of NWs with different size and shape in the same substrate (Fig. 1c). We can observe a gradient of size and shape distribution of NWs from cylindrical (lower part) to triangular (upper part) shapes. The high resolution FESEM images (outsets Fig. 1c) show triangularly shaped and almost cylindrical shaped NWs with circular cross section conceived from upper and lower portions of the substrate, respectively as indicated by dashed arrows in Figure 1c.

### 3.2 Growth Mechanism for the NWs

As described earlier in the experimental section, two independent zones in the reaction chamber were maintained at optimized temperatures of 1100 $^o$C and 900 $^o$C for the evaporation of Al and participation of Au as catalyst in the growth process, respectively (schematic Fig. 2). The Ga precursor was also kept in the zone with temperature at 900 $^o$C. At these temperatures, the metallic Al evaporated from the first zone and Ga metal evaporated from the second zone interact with molten Au catalyst particle on the Si(100) substrate to form a ternary liquid Al-Ga-Au alloy droplets. The Al-Ga-Au ternary liquid alloy absorbs atomic N liberated from decomposed NH₃ above a temperature of 400 °C.[27] The AlGaN phase is precipitated, once the amount of atomic N supersaturates in the ternary alloy on Au catalyst NPs of similar diameter at the liquid–solid interface to achieve minimum free energy of the alloy system.[16,28-31] The tentative growth process is depicted schematically in figure 2.

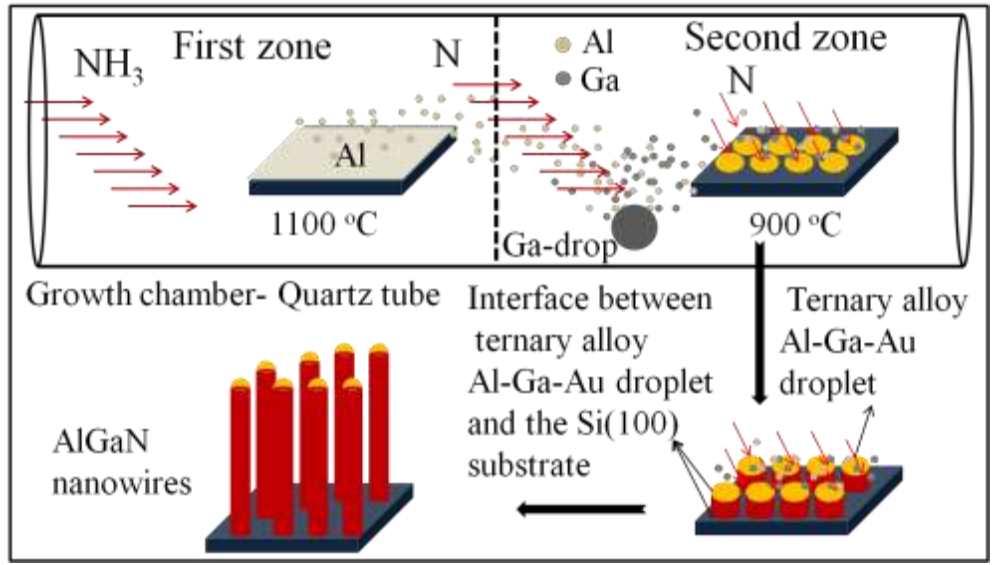

**Figure 2.** The schematic diagram for the Au catalytic VLS growth mechanism of the NWs.

The size and shape of the 1D NWs are determined by the size of the catalyst NPs used for the VLS growth purpose. Thermodynamically, the minimum radius required for a liquid Au catalyst droplet to form a eutectic alloy is,[29]

$$R_m = \frac{2V_l}{RT \ln(S)} \sigma_{lv} \quad \text{...................................................................(1)}$$

where $V_l$ is the molar volume of the droplet, $\sigma_{lv}$ is the liquid–vapor surface energy, and $S$ is the degree of supersaturation of the vapor. Thus, a higher degree of supersaturation of atomic N, which is energetically difficult to achieve in practical purpose, is required for smaller catalyst sizes. The chemical potential of the constituents in the metal–alloy catalyst becomes high, as the size of the catalyst NPs diminishes due to the Gibbs–Thompson effect:[29]

$$\Delta\mu = \frac{2\gamma}{r} \quad \text{................................................(2)}$$

where, $\Delta\mu$ is the chemical potential difference of the component species in the liquid droplet, $\gamma$ is the surface energy, and $r$ is the radius of curvature of the droplet. The Gibbs–Thompson effect predicts that, the reduction of the size of Au NPs increases the difficulty in dissolving of vapor particles in the molten Au droplet to form a eutectic alloy.[29] Hence, it will increase the degree of complexity to achieve supersaturation states that adequately promote the growth of NWs. Ostwald ripening occurs between NPs at high temperature because bigger particles are energetically favorable. Ostwald ripening is a spontaneous process such that NPs tends to transform into large particles to achieve a lower energy state, if the temperature is high enough to stimulate diffusion of the metal components. The presence Ostwald ripening leads to the development of metal catalyst droplets which may take part in the growth of 1D structures.[29,32] In the presence of Ostwald ripening, the probability of agglomeration of Au NPs at high temperature are very high for the first sample (10 min deposited) with higher Au thickness as compared to the second (5 min deposited) one. Therefore, the first sample leads to the dense distribution of the NWs with almost all NPs participating in the VLS growth mechanism. But in the case of the second sample, the number density of Au NPs is low and therefore the possibility of Ostwald ripening is very limited. So, it leads to the formation of uniformly distributed NWs with almost similar sizes and shape. The growth of NWs with a uniform radius, as observed in case of the second sample, is associated with a steady state growth in which material is transported to the particle/NW interface.[31] A constant NW diameter is possible if the incorporation of material in the vapour phase is directed exclusively on the catalyst particle where incorporation from the sidewalls is prohibited. The catalyst particle may be able to affect the size of the NW either by matching of the size of the NW or else by some mechanism involving the curvature of the particle in which strain and lattice matching play a role.[32] In case of the third sample with colloidal Au NPs, we made a gradient of Au NPs distribution in such a way that almost all the catalyst NPs participated in the process leading to formation of NWs with different diameters. The Au NPs is likely to form smaller size in the area with lower number density than that for the area with higher number density. The shape of the NWs varies from circular to triangular in cross–section for NWs grown with small to large diameter of Au NPs, respectively. Therefore, it is very important that the Au metal NPs have to be carefully aligned and separated from each other during the preparation of substrate before heating in the furnace for the growth of 1D nanostructure.

In understanding uniformity in size and controlling the density of NWs, there are two major types of VLS growth kinetics which can be envisaged.[32] The first one is single–prong growth mechanism to control the uniformity in the growth of 1D NWs. In another model, a multi–prong growth mechanism is invoked where growth density of the NWs can be understood. In the multi–prong growth, more than one NW grows from a single catalyst particle. Since NWs with comparatively smaller size are growing from a single bigger sized Au NP, the size of the individual NWs will be less than the size of the initial catalyst NP. In this case, the radius of the NW ($r_{NW}$) must be less than the radius of the catalytic Au NPs participated in the initial growth process ($r_{NP}$) (i.e. $r_{NW} < r_{NP}$). However, after the complete growth process, it is possible to observe the distribution of NWs with smaller sized Au NPs at the tip compared to the individual NW. The total process is schematically detailed in the supplementary information (Fig. S1a). In the single–prong growth, there is a one–to–one correspondence between Au NPs and NWs. In case of single–prong growth $r_{NW} \approx r_{NP}$ is usually observed but sometimes the catalyst NP is marginally larger (supplementary Fig. S1b). In multi–prong growth $r_{NW}$ is not determined directly by $r_{NP}$ but must be related to other structural factors such as the curvature of the growth interface and lattice matching between the Au catalytic NPs and the NW due to localized stress.[32] The FESEM image for the first sample (outsets of Fig. 1a) shows that the NWs are grown with triangular shape with $r_{NW} > r_{NP}$, indicating a multi-prong VLS growth mechanism. However, a single–prong growth mechanism can be envisaged for the case of the second sample with $r_{NW} \leq r_{NP}$ (outsets of Fig. 1b). The sample grown with graded Au NP number density again confirms that the lower portion of the sample follows single–prong growth mechanism with cylindrical NWs (outset of Fig. 1c). On the other hand the upper layer follows the combination of single–prong and multi–prong growth mechanism leading to the triangular shaped dense NWs (outset of Fig. 1c). Thus the size and shape of the 1D NWs are determined by the size of the catalyst NPs as well as the curvature of the Au NPs participated in the growth process. The bigger sized Au NP participated in the multi-prong mechanism leads to the triangular shaped NWs. Whereas Au NPs participated in the single-prong mechanism leading to the cylindrical NWs.

**3.3 Structural Analyses**

A typical low magnification TEM micrograph is shown (Fig. 3a) for a single NW of the mono-dispersed second sample. The SAED pattern of the NW (Fig. 3b) is indexed to wurtzite phase of single crystal AlGaN with zone axes along [0001]. HRTEM image (Fig. 3c) shows interplanar spacing of 2.76 Å corresponding to (1-100) *m*-planes of AlGaN. The inverse fast Fourier transformed (IFFT) image (Fig. 3d) of the–FFT pattern (not shown in figure) of the selected area (yellow squared) of the HRTEM image (Fig. 3c) shows an interplanar spacing of 2.76Å, which is similar as analysed from the HRTEM studies (Fig. 3c) indicating the growth direction of the NW along [1-100], which is normal to the non-polar *m*-plane of wurtzite AlGaN.

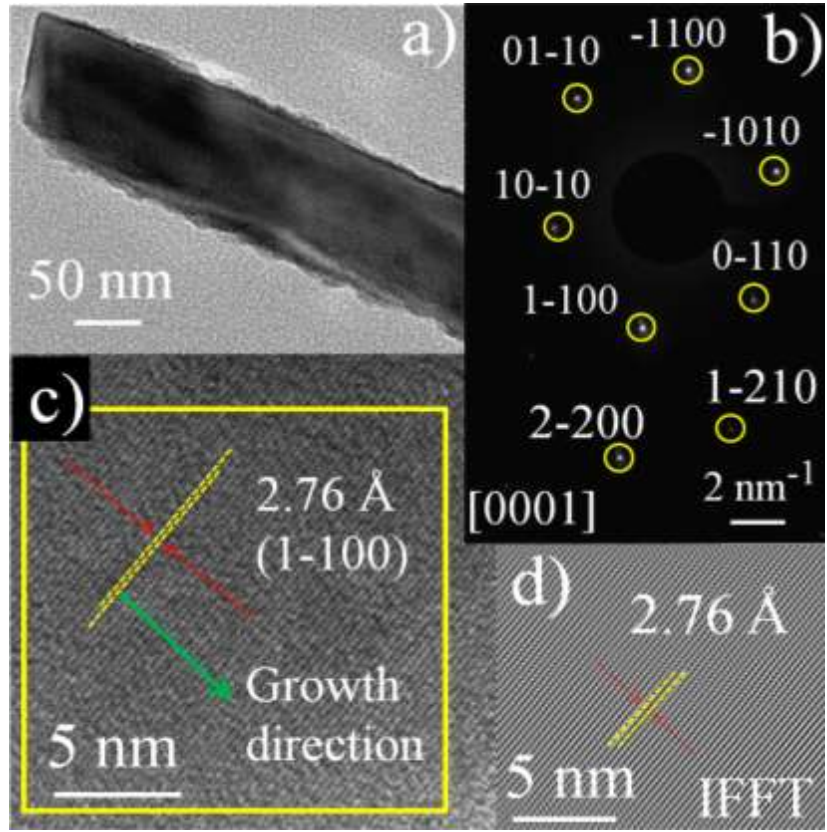

**Figure 3.** a) A low magnification TEM image of typical mono-dispersed NW. b) SAED pattern of the NW indexed to the wurtzite phase of AlGaN with zone axis along [0001]. c) HRTEM image of the AlGaN NW showing the growth direction along [1-100]. d) IFFT image of the selected (yellow squared) portion of the HRTEM image.

In order to confirm the existence of Al in the as grown AlGaN NWs, we recorded EELS spectra to detect the presence and distribution of atomic Al along the NWs. Since the percentage of Al incorporated in the AlGaN NWs is very less compared to other two components of Ga and N, the typical Al-*K* edge spectrum (Fig. 4a) collected from single AlGaN NW shows very low intensity. The Ga–$L_{2,3}$ and N–$K$ edges are shown in figures 4b and 4c, respectively. Apart from EELS study, we carried out the EFTEM for further confirmation of Al in the NWs. Figure 4d (gray color) depicts the typical zero–loss EFTEM image of AlGaN. Using this EFTEM technique, we successively imaged the NW for surface plasmon (SP), volume plasmon (VP), $L_{2,3}$ edge, and $L_1$ edge of the elemental Al. The corresponding images with golden pseudo color are represented in the figures 4e, 4f, 4g and 4h, respectively showing uniform distribution of Al. In the similar way figure 4i represents the EFTEM image (green pseudo color) for energy of 19 eV corresponding to the VP of the Ga. Finally, figure 4j represents EFTEM image with blue pseudo color for an energy 25 eV corresponds to the VP of the N. So the EFTEM imaging study of AlGaN NW is an additional confirmation for the presence of Al in sample grown in the VLS assisted CVD technique.

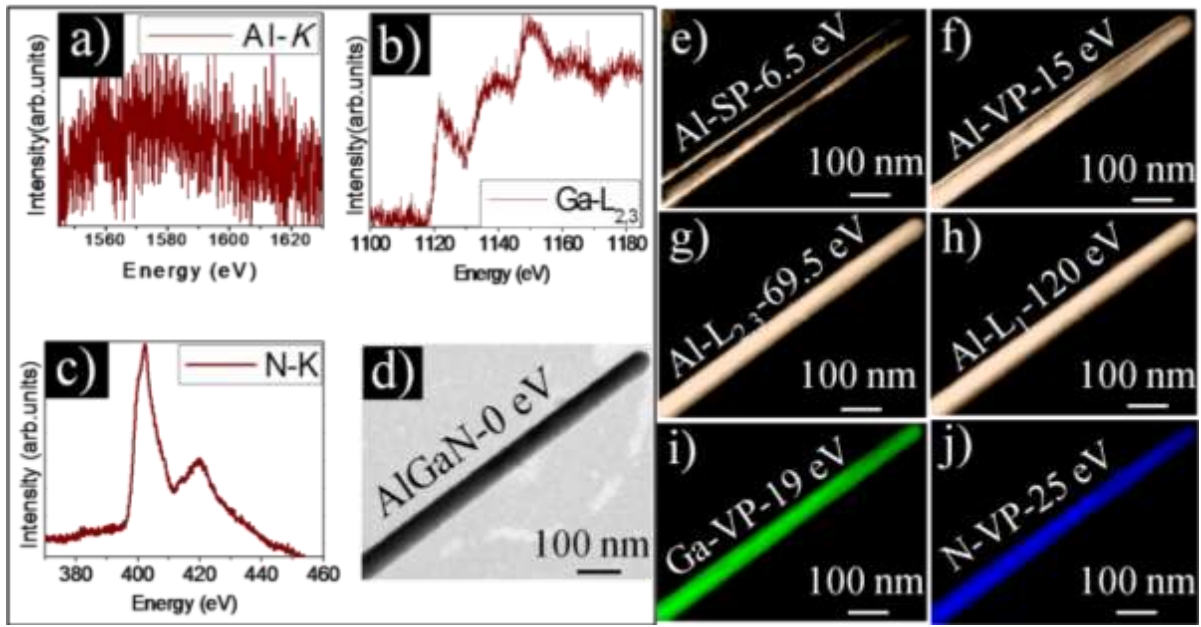

**Figure 4.** EELS spectra collected from single typical AlGaN NW for a) Al-*K* b) Ga-$L_{2,3}$ c) N-*K* edges. The EFTEM images of the same AlGaN NW for d) the zero loss. e) 6.5 eV corresponding to SP energy of Al. f) 15 eV corresponding to the VP energy of the Al g) 69.5 eV corresponding to the $L_{2,3}$ edge of Al. h) 120 eV corresponding to the $L_1$ edge of Al. i) 19 eV corresponding to the VP of Ga. j) 25 eV corresponding to the VP of N. The pseudo colors depicted in the EFTEM image chosen for describing presence of different elements of Al (golden), Ga (green) and N (blue) in the NW.

### 3.4 Optical Properties

*3.4.1 Vibrational studies for ensemble of NWs*

A typical unpolarized Raman spectrum for the ensemble of mono-dispersed AlGaN NWs is shown in the figure 5. The first–order phonon Raman scattering is caused by phonons with wave vector $k \sim 0$ ($\Gamma$ point) because of a momentum conservation rule in the light scattering process. In the hexagonal structure, group theory predicts eight sets of phonon normal modes at the $\Gamma$ point, $2A_1+2E_1+2B_1+2E_2$. Among them, one set of $A_1$ modes and one set of $E_1$ modes are both acoustic, two *B* modes are silent, while the remaining four modes, $A_1+E_1+2E_2$, are Raman-active.[33,34] Furthermore, the polar vibrational modes of $A_1$ and $E_1$ polarize the unit cell and create a long-range electrostatic field. This makes the $A_1$ and $E_1$ modes split into longitudinal optical (LO) and transverse optical (TO) modes.[34] The different modes are observed for both the samples at the peak positions of 140, 255, 420, 535, 558, 567, 668, 724 and 745 cm$^{-1}$. The corresponding symmetry assignments for observed modes are tabulated in the table 1. The simultaneous presence of GaN– $E_2^H$ centered at 567 cm$^{-1}$ along with the AlN– $E_2^H$ mode centered at 668 cm$^{-1}$ indicating the AlGaN random alloy formation. All other modes appeared in the Raman spectra correspond to the assigned modes of GaN in the wurtzite phase for the as-prepared sample. The vibrational modes for both the samples show very similar Raman symmetry modes (supplementary information in Fig. S2 showing Raman spectrum for the densely grown triangular shaped NWs).

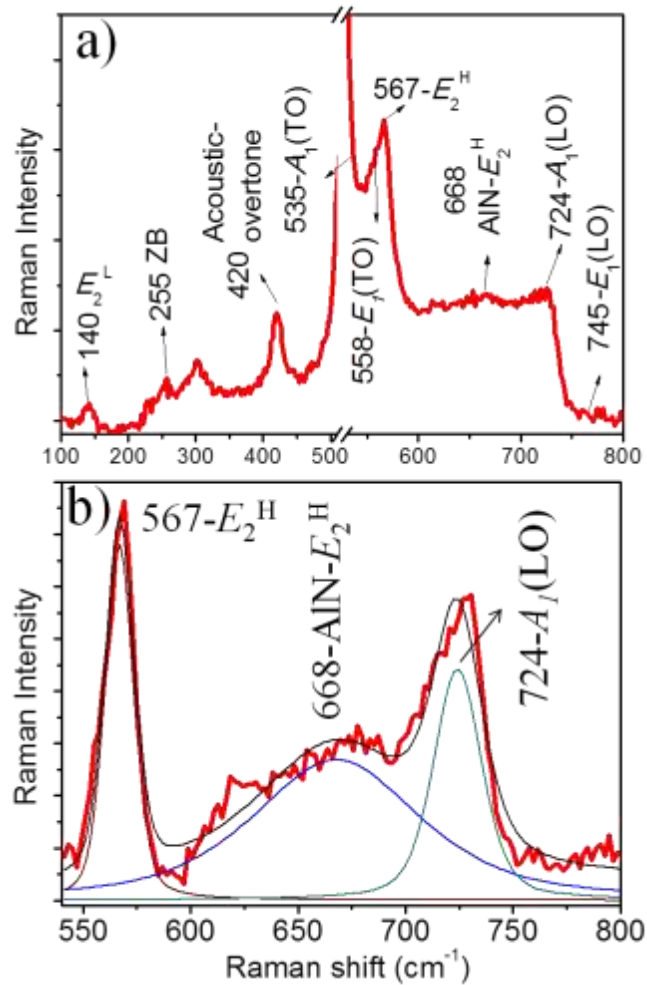

**Figure 5**. Raman spectra for a) ensemble of mono-dispersed AlGaN NWs b) Lorentzian fitted peaks in between the regions 550 - 800 cm$^{-1}$.

It is well studied that for AlGaN random alloy, both the $A_1$ (LO) and $E_1$ (LO) modes will show one-mode behavior, while the others, $A_1$(TO), $E_1$(TO), $E_2^H$ and $E_2^L$ will follow two-mode behavior.[33,35] If we recall that the phonon dispersion of wurtzite nitride which is again closely connected to that of zinc–blende nitride, LO modes tend to show one–mode behavior, while others show two–mode behavior.[35,36] The presence of $E_2^H$ modes of GaN and AlN at 567 cm$^{-1}$ and 668 cm$^{-1}$, respectively in the same Raman spectrum shows the two–mode behaviour of the random alloy that confirms again the presence of Al in the as-prepaired NWs and it is supporting stongly the EFTEM analysis carried out for the Al presence in the AlGaN NWs.

**Table 1:** Assignments of various phonon modes in the ensemble of mono-dispersed NWs

| AlGaN Raman shift (cm$^{-1}$) | Symmetry Assignments |
|---|---|
| 140 | GaN– $E_2^L$ [12,37-39] |
| 255 | GaN–Zone boundary phonon[11,39] |
| 420 | GaN–Acoustic overtone [11,12,37,39] |
| 535 | GaN–$A_1$(TO)[11,12,37,39] |
| 558 | GaN–$E_1$(TO)[12,37,38] |
| 567 | GaN–$E_2^H$ [33,38-40] |
| 668 | AlN– $E_2^H$ [10,33,41] |
| 724 | GaN–$A_1$(LO)[12,37-40] |
| 745 | GaN–$E_1$(LO)[12,37-39] (reported to be weak)[38] |

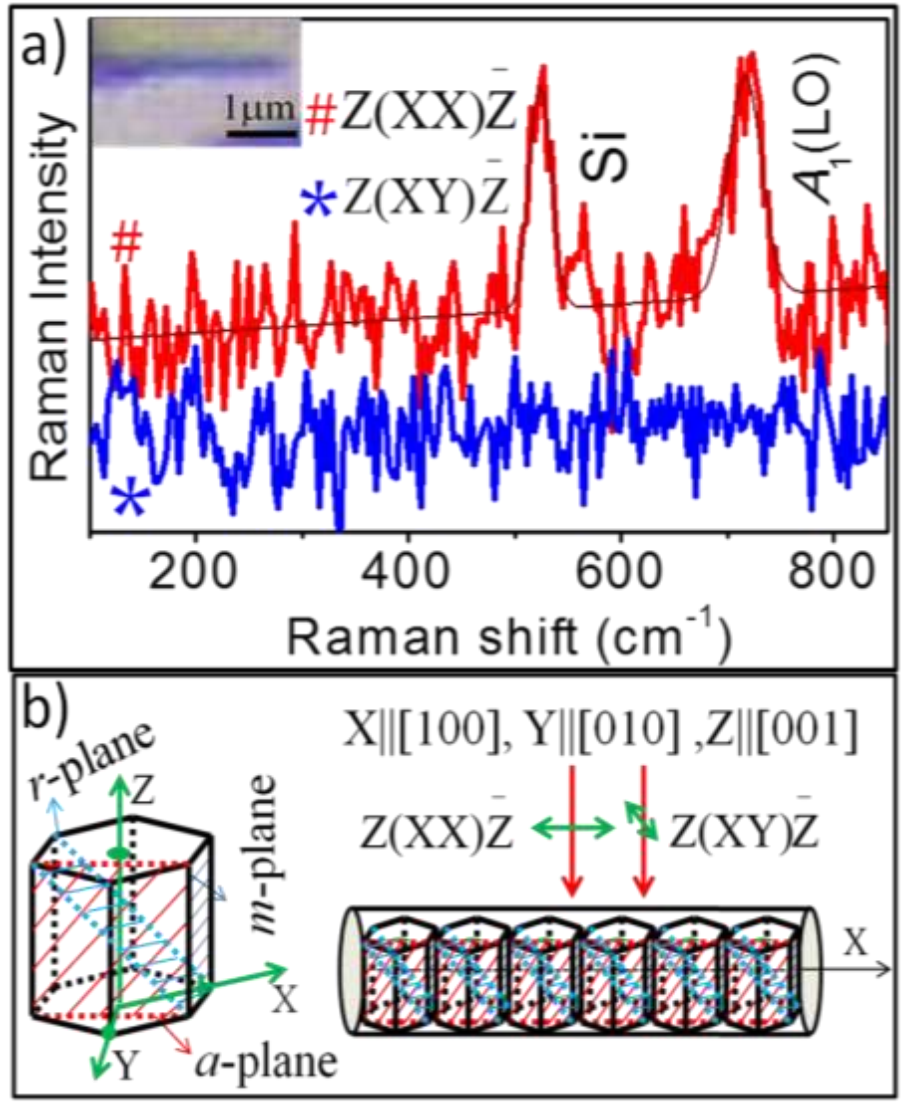

**Figure 6.** a) Polarized Raman spectra for single NW with an excitation wavelength 325nm for two different polarization configurations $Z(XX)\bar{Z}$ and $Z(XY)\bar{Z}$. The first configuration shows $A_1$(LO) mode along with the Si peak but the second configuration shows none of them. b) The schematic depiction of polarized Raman Spectra for single NW and possible growth orientation of the NW. Two different configurations used for the polarization study for the AlGaN NW and the possible stacking arrangement of unit cells in the cylindrically shaped NW.

We have also performed polarization dependent resonant Raman scattering experiments for a single AlaGaN NW for further confirmation of crystalline orientation in the mono-dispersed sample. We have to resort to resonant Raman mode using the 325 nm (3.81 eV) excitation above the band gap of GaN (3.47 eV)[13] to maximize the Raman signal for a single NW in the sub-diffraction limit by invoking Fröhlich interaction of electron-phonon coupling.[42] The polarized resonant Raman spectrum of single NW shows only one peak (Fig. 6a) centered at 722 cm$^{-1}$ which is assigned to $A_1$(LO) mode of the GaN for 325 nm excitation source. Strength of electron-phonon interaction is measured by the intensity ratios of second and first order modes which is ~ 0.6 in the present case (supplementary Fig. S3). In best of the GaN NWs, the value is reported around 0.3 –0.4.[42] The peak observed at 520 cm$^{-1}$ is because of the optical vibrational mode of the Si(100) substrate. The sample is chosen for the polarized micro–Raman spectra in such a way that single crystalline AlGaN NW is horizontally laying on the on the Si(100) substrate itself. The axis along the cylindrical NW is chosen as X direction, where as the incident and the scattered light propagation direction is considered along the Z direction. The parallel $Z(XX)\bar{Z}$ and perpendicular $Z(XY)\bar{Z}$ polarization was configured using a half wave plate and a polarizer.[12] According to the polarization selection rule in the backscattering configuration for a wurtzite [0001] GaN, (i.e., Z∥[0001]), only the $E_2^H$ phonon mode should be observable in the $Z(XY)\bar{Z}$ while both $E_2^H$ and $A_1$(LO) phonon modes should be viewed in the $Z(XX)\bar{Z}$ configuration.[12] However, it is well known that $E_2^H$ mode will not be highly active in the resonance Raman spectra. Thus only the $A_1$(LO) mode around 722 cm$^{-1}$, was observed in the $Z(XX)\bar{Z}$ configuration and no Raman peak was observed in the case of the $Z(XY)\bar{Z}$ configuration. The schematic representation of the different configurations for polarized resonance Raman spectra and the possible growth orientation of AlGaN single NW is shown in the figure 6b. Incidentally the optical mode of Si is also found to disappear in the $Z(XY)\bar{Z}$ configuration in the cubic symmetry.[43]

Hence the presence of $A_1$ (LO) mode in the $Z(XX)\bar{Z}$ configuration and the absence of it in the $Z(XY)\bar{Z}$ configuration clearly indicates that the horizontally laying NWs on the Si(100) substrate have the crystallographic growth orientations normal to the nonpolar *m*–plane [1-100] direction (schematic Fig. 6b). This is exactly matching with the information obtained from the HRTEM analysis (Fig. 3). So the polarized Raman spectroscopy is an accurate tool for understanding the crystallographic orientations even at the nanometer scale. Use of resonance Raman spectroscopy with strong electro-phonon coupling along with optical confinement are also exploited to understand it for the sub-diffraction limit of ~100 nm NW using 325 nm wavelength and N.A. of 0.5, for the first time. The optical confinement in the nanostructures of diameter ~100 nm, which is well above the quantum confinement size of 11 nm for GaN,[44] is envisaged for the polarization measurements in the dielectric contrast of ~ 5.6 for AlGaN with respect to that of 1 for the surrounding medium of Air.[45,46]

*3.4.2 PL Studies for Ensemble of NWs*

An excitation source of UV laser with 325 nm wavelengths is used to excite the NWs and the PL spectrum, recorded at room temperature (300 K), is shown for the ensemble of mono-dispersed NWs (Fig. 7). On deconvolution different peaks centered at 3.55, 3.49, 3.38, 3.29 and 3.19 eV are observed in the PL spectrum. The band edge peak centered at 3.55 eV is due to the recombination of free excitons and is found to be blue shifted from the reported value of 3.47 eV at 300K for GaN.[13,40,42] All other peaks are related to bound excitons and native defects, which are assigned to the corresponding PL peaks and tabulated in the table 2. The PL peaks with similar features were also observed for the dense triangular shaped NWs samples (supplementary Fig. S4). Along with the above peaks, we observed a significant peak centered at 3.73 eV which was assigned to $A_1$(LO) phonon mode of GaN.[13,40] The $A_1$(LO) mode is observed due to the coupling between LO phonon mode with available electron carriers, known as Fröhlich interaction, at the excitation energy above the band gap values.[42] In the presence of Fröhlich interaction a Lorentizian-Gaussian fitting for the PL peaks were best suited in our study.

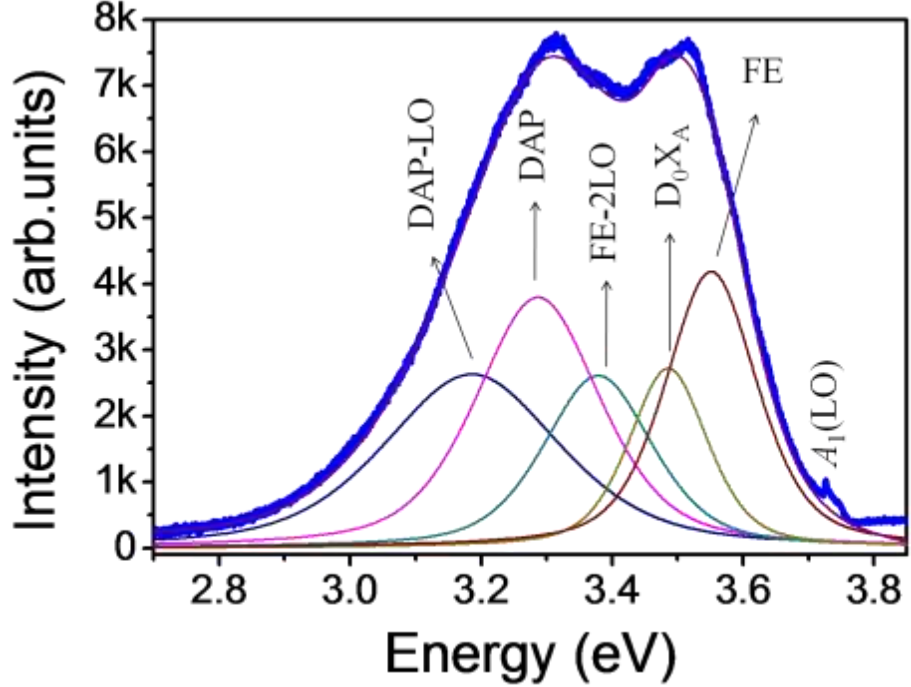

**Figure 7.** Photoluminescence spectrum observed for ensemble of mono-dispersed AlGaN NWs

**Table 2 :** Assignments of PL peaks in the ensemble of mono-dispersed NWs

| Energy (eV) | Symbols | Remarks |
|---|---|---|
| 3.53–3.55 | FE | Free Exciton emission for as prepared AlGaN NW sample. This is because of the recombination of electron–hole pair from conduction band minimum to valence–band maximum.[13,14,40,47] |
| 3.49 | $D^0X_A$ | Emission due to recombination of exciton bound to neutral donors. The possible neutral donors are the impurities (Al) those are most likely to form stable complexes with $V_N$ (i.e. Al–$V_N$).[13,14,47] |
| 3.38 | FE–2LO | Free Exciton-second order LO phonon coupling which arises due to the Fröhlich interactions.[13,40,42,47] |
| 3.29-3.33 | DAP | The recombination of the neutral donor–acceptor pair (DAP; $D^0A^0$), due to a transition from a shallow donor state of nitrogen vacancy ($V_N$) to a deep acceptor state of $V_{Ga}$. As compared to the $V_{Ga}$, the $V_{Ga}O_N$ complex is much more stable.[13-14,39-40,42,47] |
| 3.19-3.23 | DAP–LO | The phonon replica of neutral donor–acceptor pair and which was observed because of the interaction between first order $A_1$(LO) phonon with electron present in the DAP.[13,40,43] |

The three major factors affecting the blue shift in the band gap of the semiconductor nanomaterials are quantum confinement effect,[48,49] lowering temperature of the material,[50] and alloying with appropriate elements.[18-20] It is reported that the exciton Bohr radius for GaN is 11 nm,[44] so role of quantum confinement effect in elevating the band gap for as prepared AlGaN NWs to 3.55 eV with an average diameter 50-100 nm can be ruled out. The reported band gap of GaN is 3.503 eV at 0 K.[50,51] Thus the free exciton emission at 3.55 eV strongly confirms the presence of Al in the as-prepared NW sample (Fig. 7). Temperature dependent study shows blue shift of the band gap with reducing temperature (supplementary Fig. S5).

## 4. CONCLUSIONS

Ternary wurtzite phase of AlGaN nanowire is synthesised with chemical vapor deposition technique by adopting vapour-liquid-solid growth mechanism. The growth direction is found to be along [1-100], which is normal to the non-polar *m*-plane. Nanowires with different size, shape and distribution are grown, via altering the number density distribution of Au catalyst nanoparticles. The Ostwald ripening is highly probable for dense distribution of Au nanoparticles at higher temperatures which plays a key role in the size variation of nanowires. The mono-dispersed AlGaN NWs sample is envisaged in a single-prong growth mechanism. The presence of Al is confirmed by the energy-filtered transmission electron microscopic analysis. Simultaneous presence of Raman modes corresponding to GaN– $E_2^H$ at 567 cm$^{-1}$ and AlN– $E_2^H$ at 668 cm$^{-1}$ represents the two-mode behaviour of the AlGaN random alloy. A significant blue shift of Free Exciton at 3.55 eV observed in the photoluminescence spectrum for AlGaN NW as compared to the GaN band gap of 3.47 eV at 300 K confirms the presence of Al in the synthesized samples. Polarized resonance Raman spectroscopy is found to be an accurate tool for understanding the crystallographic orientations even for the sub-diffraction limit of ~100 nm NW using 325 nm wavelength. Resonance Raman spectroscopy with strong electro-phonon coupling along with optical confinement due to the dielectric contrast of nanowire with respect to that of surrounding media are used to explore it, for the first time.

**ACKNOWLEDGEMENTS**

One of us (AKS) acknowledges the Department of Atomic Energy for the financial aid. We also thank A. Das, K. K. Madapu and Venkatramana Bonu, of SND, IGCAR for their valuable suggestions and useful discussions.

**SUPPORTING INFORMATION:**

This information is available free of charge via the Internet at http://pubs.acs.org/.

**Supplementary information :**

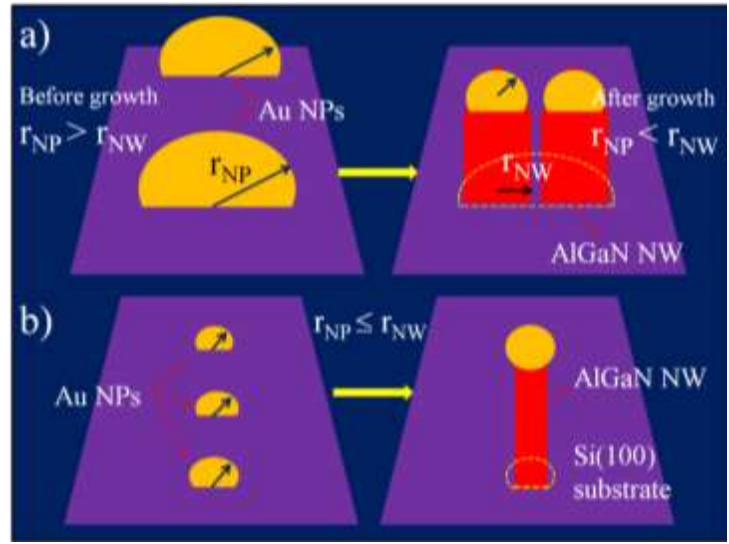

**Figure S1.** Schematic of a) multi-prong and b) single-prong growth mechanisms.

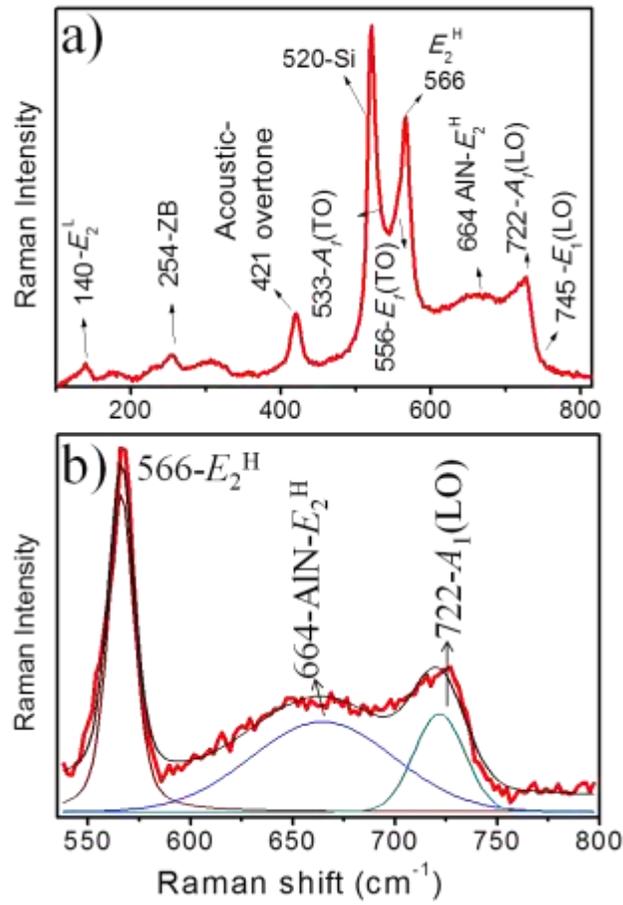

**Figure S2.** Raman spectra for a) ensemble of triangular shaped dense AlGaN NWs b) Lorentzian fitted peaks in between the regions 550 - 800 cm$^{-1}$.

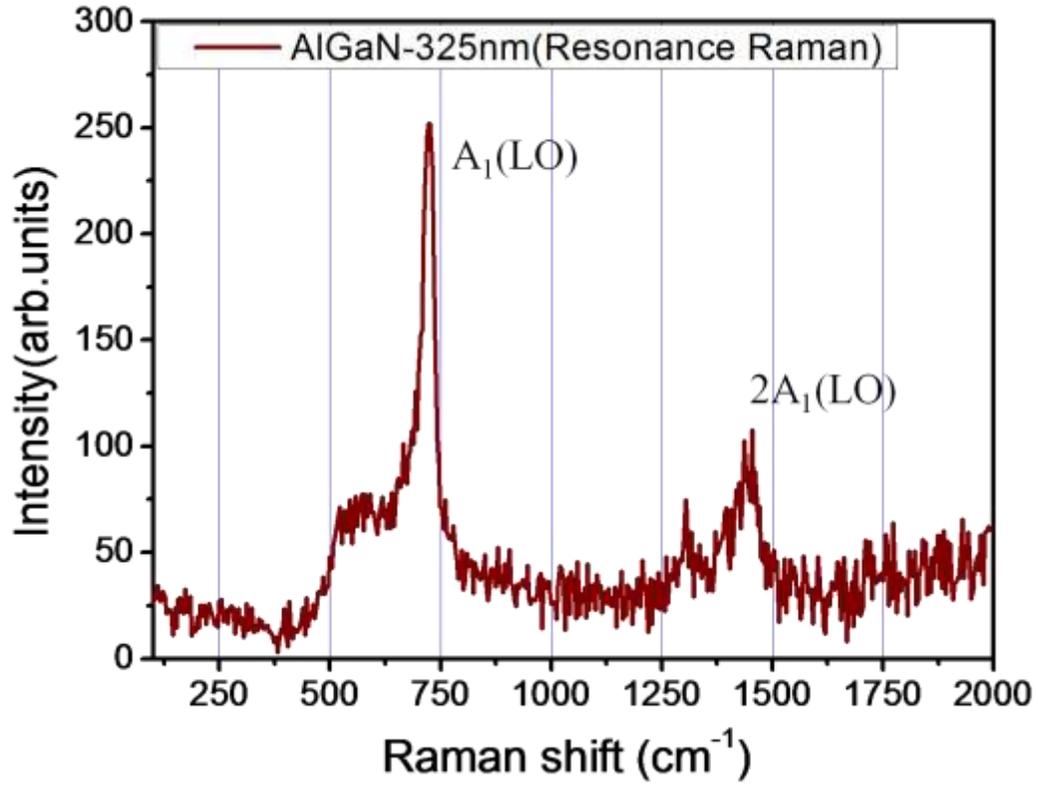

**Figure S3.** Resonance Raman spectrum observed for ensemble of AlGaN NWs with first and second order $A_1$(LO) modes. Electron-phonon coupling strength is estimated to be ~ 0.6 from the ratio of intensities of $2A_1$(LO) and $A_1$(LO).

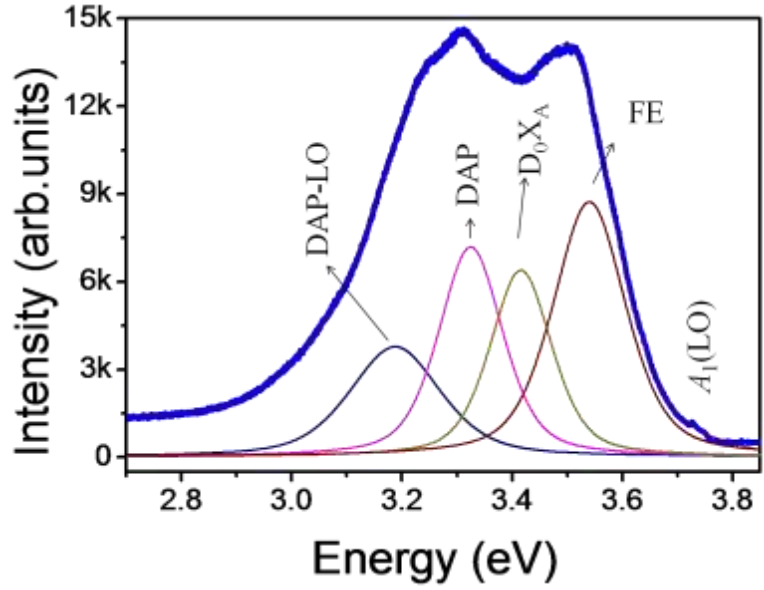

**Figure S4.** Photoluminescence spectrum observed for ensemble of triangular shaped dense AlGaN NWs

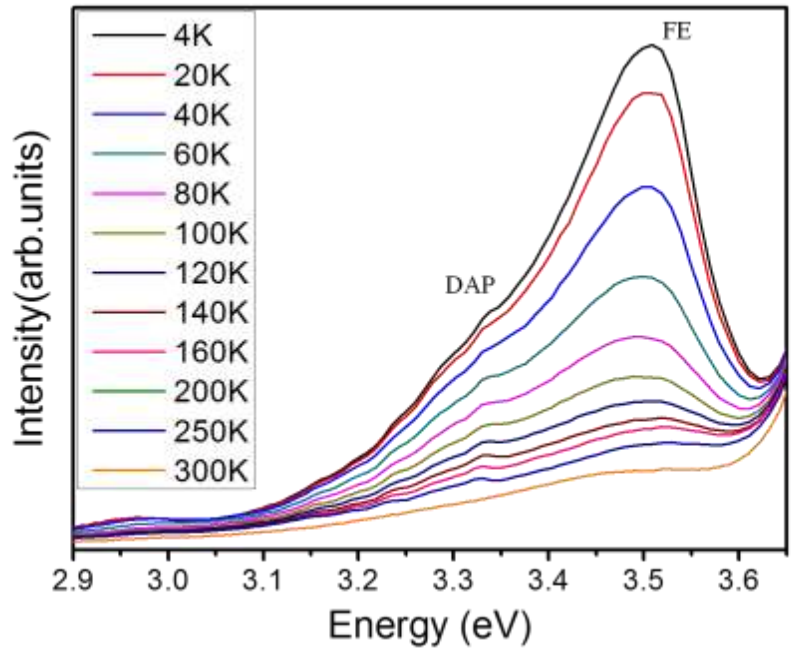

**Figure S5.** Temperature dependent PL of ensemble of triangular shaped dense AlGaN NWs